\documentclass[aps,pra,twocolumn,groupedaddress,showpacs,floatfix]{revtex4}

\usepackage[dvips]{graphicx}
\usepackage{epsfig}
\usepackage{amssymb}
\usepackage[fleqn]{amsmath}
\usepackage{color}
\usepackage{hyperref}

\begin{document}


\title{Preparation of circular Rydberg states in helium using the crossed fields method}


\author{V. Zhelyazkova and S. D. Hogan}

\affiliation{Department of Physics and Astronomy, University College London, Gower Street, London WC1E 6BT, U.K.}


\date{\today}

\begin{abstract} Helium atoms have been prepared in the circular $|n=55,\ell=54,m_{\ell}=+54\rangle$ Rydberg state using the crossed electric and magnetic fields method. The atoms, initially travelling in pulsed supersonic beams, were photoexcited from the metastable $1s2s\,^3S_1$ level to the outermost, $m_{\ell}=0$ Rydberg-Stark state with $n=55$ in the presence of a strong electric field and weak perpendicular magnetic field. Following excitation, the electric field was adiabatically switched off causing the atoms to evolve into the circular state with $m_{\ell}=+54$ defined with respect to the magnetic field quantization axis. The circular states were detected by ramped electric field ionization along the magnetic field axis. The dependence of the circular state production efficiency on the strength of the excitation electric field, and the electric-field switch-off time was studied, and microwave spectroscopy of the circular-to-circular $|55,54,+54\rangle\rightarrow|56,55,+55\rangle$ transition at $\sim38.5$~GHz was performed. 
\end{abstract}


\maketitle

\section{Introduction}
Until recently, experimental approaches to quantum information processing and quantum computation have typically centered around the use of individual quantum systems, e.g., atoms or ions, in which coherence times as long as 50~s can be achieved and quantum gates can be implemented with near-unit fidelity~\cite{harty14,monroe02} but which are challenging to scale to incorporate large numbers of qubits~\cite{jurcevic14}; or solid state devices (e.g., Ref.~\cite{blais04}) which are inherently scalable but exhibit limited coherence times~\cite{railly15}. However, attempts are now being made to investigate hybrid architectures for quantum information processing in which quantum systems with complementary properties are combined~\cite{wallquist09,xiang13}. One such hybrid architecture, suitable for operation at microwave frequencies, involves combining gas-phase Rydberg atoms and solid-state superconducting microwave circuits, with the coupling between the two systems achieved via chip-based co-planar microwave resonators~\cite{hogan12,rabl06}. Such an architecture can be considered a hybrid between traditional three-dimensional microwave cavity quantum electrodynamics (QED)~\cite{haroche13}, and circuit QED~\cite{wallraff04}.

Rydberg atoms are well-suited to such hybrid approaches to microwave cavity QED because they possess large transition dipole moments at microwave frequencies. In particular, atoms in circular Rydberg states, i.e., states in which the excited electron has maximum orbital angular momentum, $\ell$, and azimuthal, $m_{\ell}$, quantum numbers ($|m_{\ell}|=\ell=n-1$), are uniquely suitable for such experiments. These states (i) are characterized by long radiative lifetimes, e.g., the lifetime of the circular $|n,\ell,m_{\ell}\rangle = |55,54,+54\rangle$ state in helium is 46~ms, (ii) if coupled via $\Delta n=+1$, $\Delta m_{\ell}=+1$ microwave transitions, which lie below 50~GHz for $n>50$ and are therefore compatible with readily available microwave sources and circuits, form quasi two-level systems with large transition dipole moments, e.g., $\langle 56,55,+55|\hat{\mu}|55,54,+54\rangle=2148\,ea_0$, and (iii) exhibit no first-order Stark shifts, minimizing their sensitivity to stray electric fields emanating from surfaces~\cite{hogan12}, and extending their coherence times~\cite{avigliano14}.

Atoms in Rydberg states have a long history in microwave cavity QED experiments~\cite{haroche13}, having been exploited in the single-atom maser~\cite{meschede85}, in tests of the Jaynes-Cummings model~\cite{rempe87}, in the first demonstration of the production of Bell states using massive particles~\cite{hagley97}, and in non-demolition measurements of single photons in three-dimensional cavities~\cite{gleyzes07}. With the prospect of exploiting circular Rydberg states in hybrid cavity QED, controlled phase gates involving atoms in these states have recently been analyzed~\cite{xia13}, as has the use of non-resonant microwave dressing to suppress their dc-electric-field sensitivity~\cite{ni15}.

Because of their high-$\ell$ character, circular Rydberg states cannot be directly photoexcited from an atomic or molecular ground state as laser excitation typically only populates low-$\ell$ Rydberg states. However, two main methods for converting low-$\ell$ Rydberg states into circular states have been developed. These are the adiabatic microwave transfer method, in which a sequential increase of $|m_{\ell}|$ in a series of microwave transitions converts optically accessible low-$|m_{\ell}|$ states into circular states~\cite{hulet83,liang86,cheng94}, and the crossed electric and magnetic fields method~\cite{delande88,hare88}. In the crossed fields method atoms are excited to the $m_{\ell}=0$ Rydberg-Stark state with maximal energy for a particular value of $n$, in the presence of a dc electric field (i.e., the outermost low-field-seeking Stark state) and a weak perpendicular magnetic field. Following excitation, the electric field is adiabatically switched off, leaving the atoms in the circular state in the magnetic field. The crossed fields method is in principle applicable to any atomic or molecular species, whereas the microwave transfer method can only be easily applied to lighter elements~\cite{delande88}.

In the past the crossed fields method has been successfully used to prepare circular Rydberg states, with values of $n$ ranging from 20 to 70 and near 100\% efficiency after initial photoexcitation, in lithium~\cite{hare88}, rubidium~\cite{brecha93,anderson13}, hydrogen~\cite{lutwak97}, and barium~\cite{cheret89}. Here we demonstrate the preparation of circular Rydberg states in helium using the crossed fields method. Helium atoms are particularly attractive for use in hybrid cavity QED experiments because (i) they cause fewer detrimental effects if adsorbed on cryogenic surfaces than alkali metal atoms~\cite{hattermann12,thiele14}, (ii) the radiation required for two-color two-photon excitation of high Rydberg states from the metastable $1s2s\,^3S_1$ level lies in convenient diode laser wavelength ranges, (iii) they can be laser cooled in the metastable triplet $1s2s\,^3S_1$ level~\cite{rooijakkers96,chang14}, and (iv) they are light and in Rydberg-Stark states can be easily transported and trapped above chip-based electrical transmission lines and resonators using inhomogeneous electric fields~\cite{lancuba13,lancuba14,lancuba16}.

In the following we first give, in Section II, an overview of the apparatus and methods used in the experiments. In Section III we provide the theoretical background and experimental evidence for the preparation of circular Ryd\-berg states in helium using the crossed fields methods. Finally, in Section IV, the results of microwave spectroscopy experiments are presented and discussed. 

\section{Experimental set-up}
\label{sec1}
\begin{figure}[!h]
\includegraphics[width=0.5\textwidth]{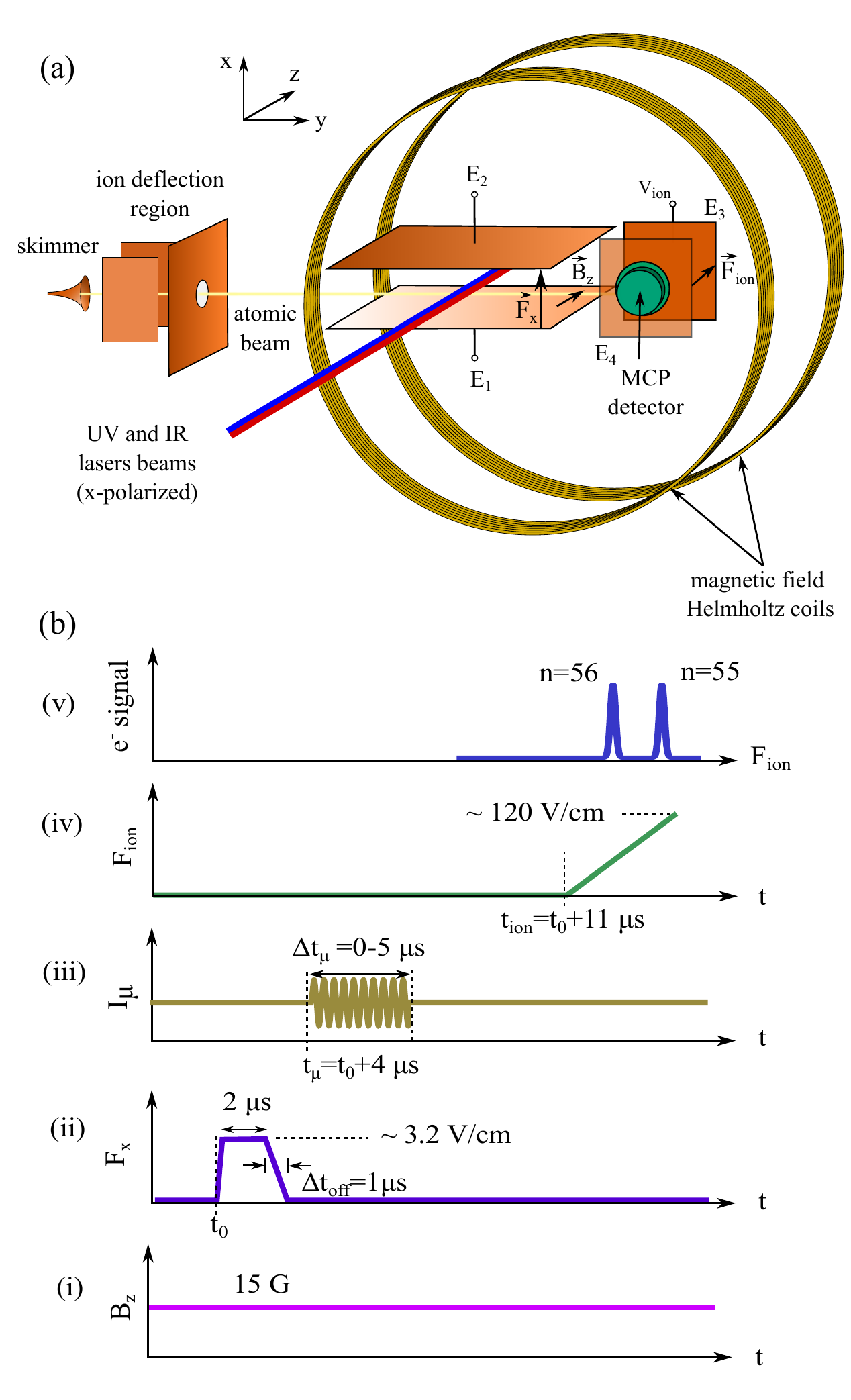}
\caption{(a) Schematic diagram of the experimental apparatus. (b) Timing of the electric field and microwave pulses employed in the experiments (see text for details).}
\label{fig1}
\end{figure}

A schematic diagram of the experimental apparatus is displayed in Fig.~\ref{fig1}. Metastable helium atoms are produced in a pulsed supersonic source, operated at a repetition rate of 50~Hz. An electric discharge~\cite{halfmann00}, generated by applying $+250$~V to an anode located at the exit of a pulsed valve populates the metastable $1s2s\,^3S_1$ level. To optimize the shot-to-shot stability, the discharge is seeded with electrons emitted by a tungsten filament positioned $\sim2$~cm downstream from the anode. The mean longitudinal speed of the beam is $\sim2000$~m\,s$^{-1}$. After passing through a skimmer with a diameter of 2~mm, located 5~cm from the valve, the atoms travel between a set of electrodes that deflect any stray ions produced in the discharge. Approximately 20~cm from the valve, the atomic beam is intersected perpendicularly by continuous IR (Toptica TA Pro) and UV (Toptica DL-SHG Pro) laser beams. The UV laser is frequency-stabilized to drive the $1s2s\,^3S_1\rightarrow 1s3p\,^3P_2$ transition, at a wavelength of $\lambda_{\mathrm{UV}}=388.975$~nm, and typical intensity of 1~W\,cm$^{-2}$. The IR laser, $\lambda_{\mathrm{IR}}\simeq786.755$~nm and intensity 2000~W\,cm$^{-2}$, is scanned or stabilized for excitation of $ns/nd$ Rydberg states close to $n=55$.

The laser beams intersect the atomic beam between two square 70~mm~$\times$~70~mm copper plates [labelled $E_1$ and $E_2$ in Fig.~\ref{fig1}(a)], separated in the $x$ dimension by 8.4~mm. When the beam of metastable helium atoms reaches the point of intersection with the laser beams, a pulsed potential is applied to $E_2$ to generate a selected excitation electric field in the interaction region, bringing the IR laser into resonance with the transition to the Rydberg state. Both laser beams are linearly polarized parallel to this excitation electric field. Additional stray electric fields, emanating from the copper plates in the excitation region, are cancelled to $\pm1$~mV\,cm$^{-1}$. The rise time of the potential applied to $E_2$ is $200$~ns, after which it is held constant for 2~$\mu$s [Fig.~\ref{fig1}(b-ii)]. Considering the average longitudinal speed of the atomic beam, a 4~mm long bunch of Rydberg atoms is therefore excited during the activation time of this excitation electric field pulse. Throughout the excitation region a magnetic field of $\sim15$~G is generated by a pair of Helmholtz coils wound around the outside of the vacuum chamber [Fig.~\ref{fig1}(b-i)]. The magnetic field is oriented perpendicular to the excitation electric field to a precision on the order of $\pm0.1^{\circ}$. After excitation, the electric field is switched off allowing the excited atoms to evolve into the pure magnetic field.

After this preparation phase, the Rydberg atoms can be exposed to microwave fields to drive $\Delta n = +1$ transitions at $\sim38.5$~GHz. To achieve this pulsed microwave radiation is generated using an Agilent Technologies PSG Analog Signal Generator (E8257D) the output of which is coupled into the interaction region in free space through a 15~mm-diameter quartz window. The microwave pulses are applied 1.5~$\mu$s after Rydberg state photoexcitation, for time durations up to 5~$\mu$s [Fig.~\ref{fig1}(b-iii)]. The microwave intensity in the interaction region is calibrated by measurements of Rabi oscillations in Rydberg-Rydberg transitions.

After exiting the interaction region, the Rydberg atoms fly into the detection region of the apparatus, where they evolve adiabatically into an electric field ($\sim0.5$~V\,cm$^{-1}$) oriented parallel to the background magnetic field~\cite{cheret89}. When the atoms arrive in front of a microchannel plate (MCP) detector, a pulsed electric field is ramped from 0 to $-120$~V\,cm$^{-1}$ in 5~$\mu$s by applying a time-dependent potential to electrode $E_3$. This field ionizes the Rydberg atoms in the direction parallel to the magnetic field and accelerates the resulting electrons toward the MCP. The electron current on the MCP is then collected as a measure of the Rydberg state population. 

\section{Circular state preparation}
\label{sec:2}
The energy level structure of the triplet Rydberg states of helium in the presence of perpendicularly crossed electric and magnetic fields can be calculated by determining the eigenvalues of the Hamiltonian matrix 
\begin{eqnarray}
\hat{H} &=& \hat{H}_0 + \hat{H}_{\mathrm{Z},L} + \hat{H}_{\mathrm{Z},S}+ \hat{H}_{\mathrm{S}},
\label{eqHamiltonian}
\end{eqnarray} 
where $\hat{H}_0$ is the field-free Hamiltonian, $\hat{H}_{\mathrm{Z},L}$ and $\hat{H}_{\mathrm{Z},S}$ represent the Zeeman interaction of the orbital magnetic moment of the Rydberg electron, and the total spin magnetic moment of the atom with the external magnetic field, respectively, and $\hat{H}_{\mathrm{S}}$ represents the Stark interaction with the electric field. For a magnetic field $\vec{B}=(0,0,B_z)$ and an electric field $\vec{F}=(F_x,0,0)$ the Hamiltonian in the weak magnetic field limit, i.e., when $ea_0^2Bn^{7/2}/\hbar\ll1$ and the diamagnetic interaction can be neglected~\cite{garstang77}, takes the form
\begin{eqnarray}
\hat{H}&=&\hat{H}_0+\frac{\mu_B}{\hbar}\hat{\vec{B}}\cdot\hat{\vec{L}}+\frac{\mu_B}{\hbar}\hat{\vec{B}}\cdot\hat{\vec{S}}+e\hat{\vec{F}}\cdot\hat{\vec{r}}\nonumber\\
&=&\hat{H}_0+\frac{\mu_B}{\hbar}B_z\hat{L}_z+\frac{\mu_B}{\hbar}B_z\hat{S}_z+eF_x\hat{x}.
\label{Hnew}
\end{eqnarray} 
Choosing an $|n,\ell,m_{\ell}\rangle$ basis in which to construct the Hamiltonian matrix, $\hat{H}_0$ is diagonal with the energy of each sublevel given by the Rydberg formula. The quantum defects, $\delta_{n,\ell}$, for the low-$\ell$ states of interest with $n=55$ are listed in Table~\ref{table1}~\cite{drake99a}.
\begin{table}[!h]
   \begin{tabular}{c}
	\hline\hline
      \hspace*{1cm}Quantum defect ($\delta_{n,\ell}$)\hspace*{1cm} \\
	 \hline
	 \hline
     $\delta_{55,0}$=0.296669 \\
     $\delta_{55,1}$= 0.068354 \\
	   $\delta_{55,2}$= 0.002889 \\
	  $\delta_{55,3}$= 0.000447\\
	  $\delta_{55,4}$= 0.000127 \\
	  $\delta_{55,5}$= 0.000049 \\
	\hline	
	\end{tabular}
	\caption{Quantum defects of the triplet Rydberg states of helium with $n=55$ and $\ell\leq5$~\cite{drake99a}.}
\label{table1}
\end{table}
In the $|n,\ell,m_{\ell}\rangle$ basis, the interaction of the orbital magnetic moment of the Rydberg electron with the magnetic field is also represented by a diagonal matrix with elements $\langle n,\ell, m_{\ell}|\hat{H}_{\mathrm{Z},L}|n, \ell, m_{\ell}\rangle=M_LB_z\mu_B=m_{\ell}B_z\mu_B$ ($M_L$ is the projection of the total orbital angular momentum onto the magnetic field axis). The interaction of the electron spin magnetic moment with the magnetic field, if treated as a perturbation, leads to a global energy shift $\Delta E_{\mathrm{Z},S}=M_SB_z\mu_{B}$, and the corresponding matrix is therefore also diagonal in the $|n,\ell, m_{\ell}\rangle$ basis for each value of $M_S$ ($M_S$ is the projection of the total spin angular momentum onto the magnetic field axis). In the experiments described here the linearly polarized UV laser radiation was frequency-stabilized to drive the $\Delta M_J=+1$ Zeeman component of the $1s2s\;^3S_1\rightarrow1s3p\;^3P_2$ transition ($J$ and $M_J$ are the total electronic angular momentum and corresponding azimuthal quantum numbers, respectively). Consequently, the intermediate state had predominantly $M_J=+2$, and hence $M_S=+1$, character. Subsequent Rydberg state photoexcitation with IR laser radiation of the same linear polarization as that of the UV radiation preserved this predominantly $M_S=+1$ character in the excited states.

\begin{figure}[!h]
\includegraphics[width=0.46\textwidth]{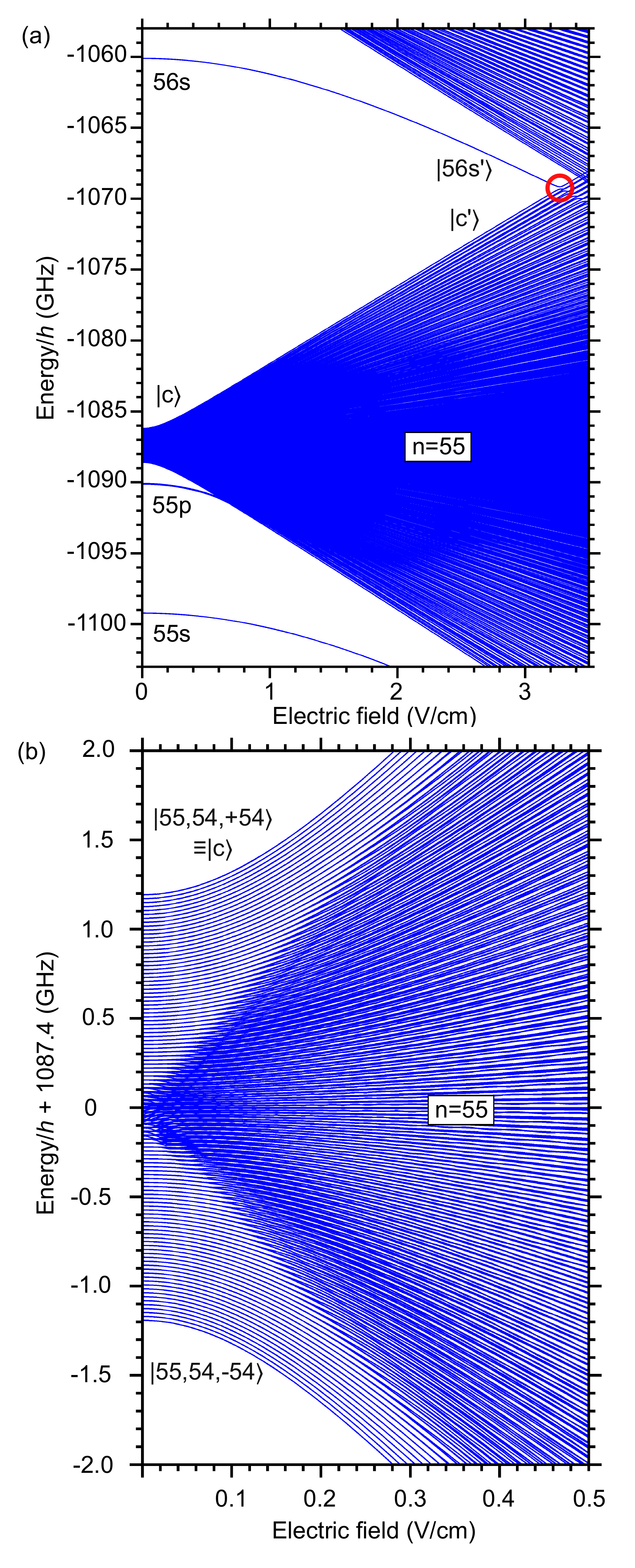}
\caption{Calculated Stark map of the triplet Rydberg states in helium with $n=55$ and $n=56$ (a) in strong electric fields up to the Inglis-Teller limit, and (b) in weak electric fields. In each calculation $B_z=15.776$~G. The red circle in (a) indicates the region of the Stark map encompassed in Fig.~\ref{fig3}.} 
\label{fig2}
\end{figure}

The Hamiltonian, $\hat{H}_{\mathrm{S}}$, associated with the interaction of the atom with the perpendicular electric field can be expanded in spherical polar coordinates in terms of the spherical harmonic functions, $Y_{\ell,m_{\ell}}(\theta,\phi)$, as
\begin{eqnarray}    
\hat{H}_{\mathrm{S}}&=&eF_x\hat{x} \nonumber\\
&=&eF_xr\sin\theta\cos\phi \nonumber\\
&=&eF_xr\sqrt{\frac{2\pi}{3}}[Y_{1,-1}(\theta,\phi)-Y_{1,1}(\theta,\phi)].
\label{eq:SH}
\end{eqnarray}
This gives rise to matrix elements which can be separated into a radial and angular integrals such that
\begin{eqnarray}
\langle n',\ell',m_{\ell}'|\hat{H}_S|n,\ell,m_{\ell}\rangle =  \qquad \qquad  \qquad \qquad \nonumber\\
=eF_x\langle n',\ell'|r|n,\ell\rangle\langle\ell,'m_{\ell}'|\sin\theta\cos\phi|\ell, m_{\ell}\rangle.
\label{HSme}
\end{eqnarray} 
The radial integrals can be calculated using the Numerov method~\cite{gallagher94,zimmerman79}. Making use of the properties of the spherical harmonic functions, the angular integrals can be written in a concise form as
\begin{eqnarray}
\langle\ell+1,m_{\ell}+1|\sin\theta\cos\phi|\ell, m_{\ell}\rangle = \qquad\qquad \nonumber\\ 
\frac{1}{2}\times(-1)^{m_{\ell}-2\ell}\sqrt{\frac{(\ell+m_{\ell}+1)(\ell+m_{\ell}+2)}{(2\ell+1)(2\ell+3)}},\\
\nonumber\\
\langle\ell-1,m_{\ell}+1|\sin\theta\cos\phi|\ell, m_{\ell}\rangle = \qquad\qquad \nonumber\\
-\frac{1}{2}\times(-1)^{-m_{\ell}+2\ell}\sqrt{\frac{(\ell-m_{\ell}-1)(\ell-m_{\ell})}{(2\ell-1)(2\ell+1)}},\\
\nonumber\\
\langle\ell+1,m_{\ell}-1|\sin\theta\cos\phi|\ell, m_{\ell}\rangle = \qquad\qquad \nonumber\\
\frac{1}{2}\times(-1)^{m_{\ell}-2\ell}\sqrt{\frac{(\ell-m_{\ell}+1)(\ell-m_{\ell}+2)}{(2\ell+1)(2\ell+3)}},\\
\nonumber\\
\langle\ell-1,m_{\ell}-1|\sin\theta\cos\phi|\ell, m_{\ell}\rangle = \qquad\qquad \nonumber \\
-\frac{1}{2}\times (-1)^{-m_{\ell}+2\ell}\sqrt{\frac{(\ell+m_{\ell}-1)(\ell+m_{\ell})}{(2\ell-1)(2\ell+1)}}.
\end{eqnarray}
After constructing the Hamiltonian matrix in Eq.~(\ref{Hnew}), the Stark energy level diagram in Fig.~\ref{fig2} was calculated by evaluating the eigenvalues in each electric field of interest. In these calculations $B_z=15.776$~G to match the experimental conditions (see Section IV). To obtain convergence over the range of energies encompassed in Fig.~\ref{fig2}, basis states with values of $n$ ranging from 54 to 57 were used. The Hamiltonian matrix therefore had dimensions of $12326\times12326$.
 
From the expanded Stark map in Fig.~\ref{fig2}(b) it can be seen that the magnetic field lifts the degeneracy of the $m_{\ell}$ sublevels. In zero electric field the Zeeman splitting between each sublevel, which corresponds to the Larmor frequency, $\nu_{\mathrm{Z}}$, is $\nu_{\mathrm{Z}}=\mu_BB_z/h\simeq22$~MHz. However, under these conditions the outermost, circular state $|\mathrm{c}\rangle\equiv|55,54,+54\rangle$ cannot be accessed directly by laser excitation from the intermediate $1s3p\;^3P_2$ level because of the restrictions imposed by the selection rules for electric dipole transitions. To provide this state with some low-$|m_{\ell}|$ character and make it accessible by laser photoexcitation, a strong perpendicular electric field is applied. This causes the state $|\mathrm{c}\rangle$ to evolve into the outermost $n=55$ Stark state in the electric field, $|\mathrm{c'}\rangle$ in Fig.~\ref{fig2}(a). As can be seen from Eq.~\ref{eq:SH}, the perpendicular electric field mixes states with $\Delta m_{\ell}=\pm1$. When the Stark energy shifts exceed the Zeeman splitting, the state $|\mathrm{c}\rangle$ has evolved into the $m_{\ell}=0$ state with respect to the electric field quantization axis and can therefore be directly laser photoexcited from the $1s3p\;^3P_2$ level. After photoexcitation in the presence of such an electric field, the pure circular state, $|\mathrm{c}\rangle$, defined with respect to the magnetic field quantization axis, can be prepared by adiabatically switching off the electric field.

\begin{figure}[!h]
\includegraphics[width=0.5\textwidth]{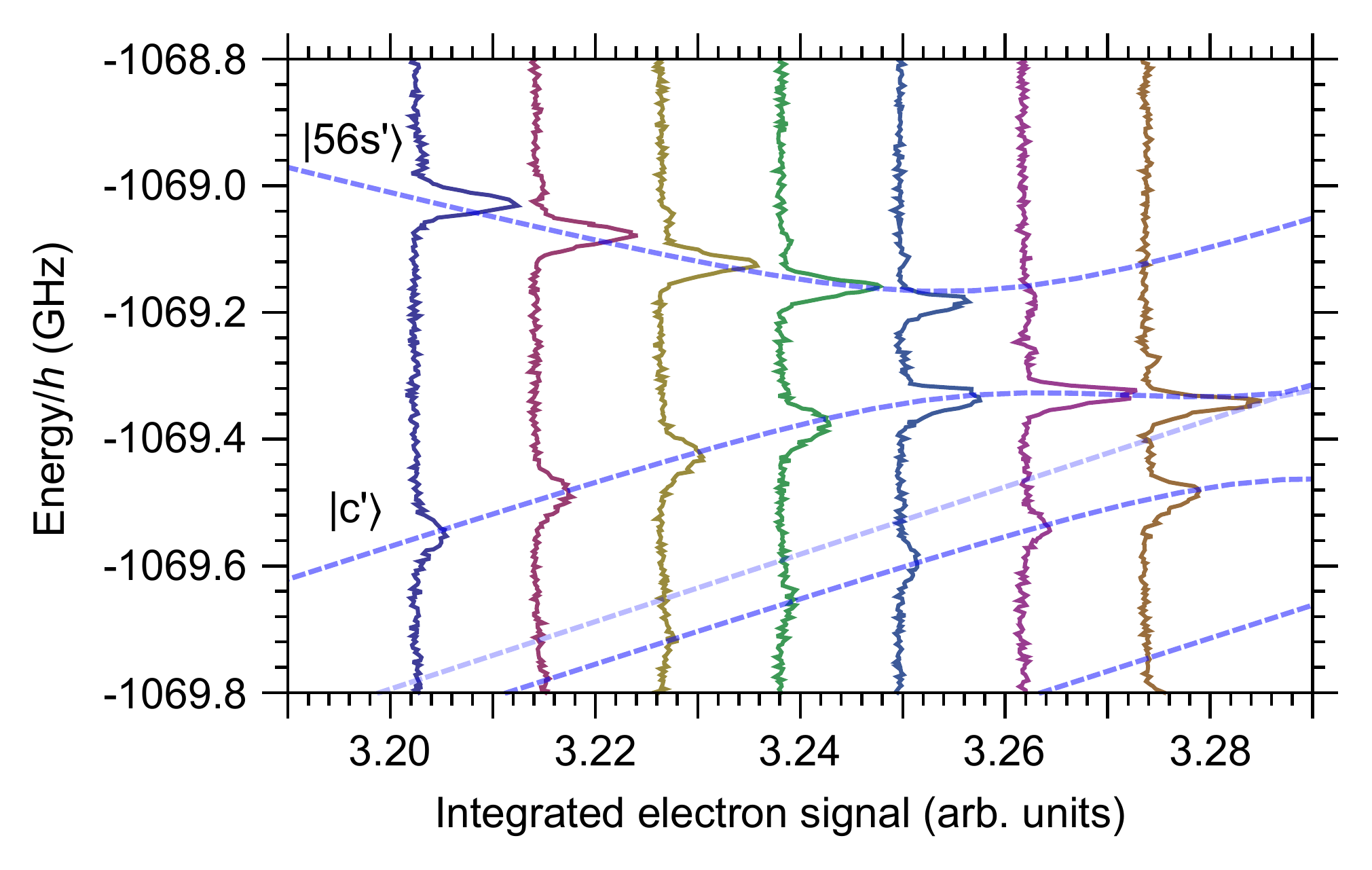}
\caption{Experimentally recorded laser photoexcitation spectra near the Inglis-Teller limit between $n=55$ and $n=56$, where the states $|\mathrm{c'}\rangle$ and $|56\mathrm{s'}\rangle$ undergo an avoided crossing. The spectra were recorded with $B_z=15.776$~G and electric fields ranging from $F_x = 3.202$ to 3.274~V\,cm$^{-1}$ as indicated by the horizontal offset in each case. The calculated electric field dependence of the energy level structure is indicated by the dashed curves.}
\label{fig3}
\end{figure}

The first step in the circular state preparation process involves excitation of the state $|\mathrm{c'}\rangle$ in the electric field. In the triplet Rydberg states of helium this state obtains significant $\ell=0$ character in electric fields close to the Inglis-Teller limit~\cite{gallagher94}, where Stark states with values of $n$ that differ by $\pm1$ first cross in the Stark map [circle in Fig.~\ref{fig2}(a)]. This allows efficient laser excitation from the $1s3p\;^3P_2$ level in such fields. Laser photoexcitation spectra recorded in this spectral region, where the state $|56\mathrm{s'}\rangle$, that evolves adiabatically to the 56s state in zero electric field [See Fig.~\ref{fig2}(a)], and the state $|\mathrm{c'}\rangle$ cross most closely, are displayed in Fig.~\ref{fig3}. In this figure the experimentally recorded spectra are overlaid with the calculated electric field dependence of the energy level structure (dashed curves), over the range of electric fields indicated on the horizontal axis. In electric fields below this avoided crossing (e.g., when $F_x<3.25$~V\,cm$^{-1}$) the spectral intensity of the transition to the state $|56\mathrm{s'}\rangle$ is greater than that to the state $|\mathrm{c'}\rangle$ required for circular state preparation. However, as the electric field strength is increased to the position of the first avoided crossing, the spectral intensity is transferred to the latter. Consequently, photoexcitation of the state $|\mathrm{c'}\rangle$ is significantly enhanced at this position in the Stark map. All subsequent measurements discussed below were therefore performed in an excitation electric field of $F_x=3.262$~V\,cm$^{-1}$.

Following photoexcitation of the state $|\mathrm{c'}\rangle$, the circular state preparation process is completed by adiabatically switching the excitation electric field to zero. The adiabaticity condition for this process requires that
\begin{eqnarray}
\frac{\mathrm{d}\nu_{\mathrm{S}}}{\mathrm{d}t} &\ll& \frac{(\nu_{\mathrm{Z}}^2+\nu_{\mathrm{S}}^2)^{3/2}}{\nu_{\mathrm{Z}}},
\end{eqnarray}
where $\nu_{\mathrm{S}}=\frac{3}{2}nF_xea_0/h$ corresponds to the energy interval between adjacent Stark states~\cite{delande88}. For an excitation electric field $F_x=3.262$~V\,cm$^{-1}$, and a magnetic field $B_z=15.776$~G, $\nu_{\mathrm{S}}=344$~MHz and $\nu_{\mathrm{Z}}=22$~MHz. Therefore if the electric field is switched to zero on time scales on the order of 1~$\mu$s the adiabaticity condition is satisfied, and close to 100\% circular state conversion efficiency is expected~\cite{delande88}. 

\begin{figure}[!h]
\includegraphics[width=0.45\textwidth]{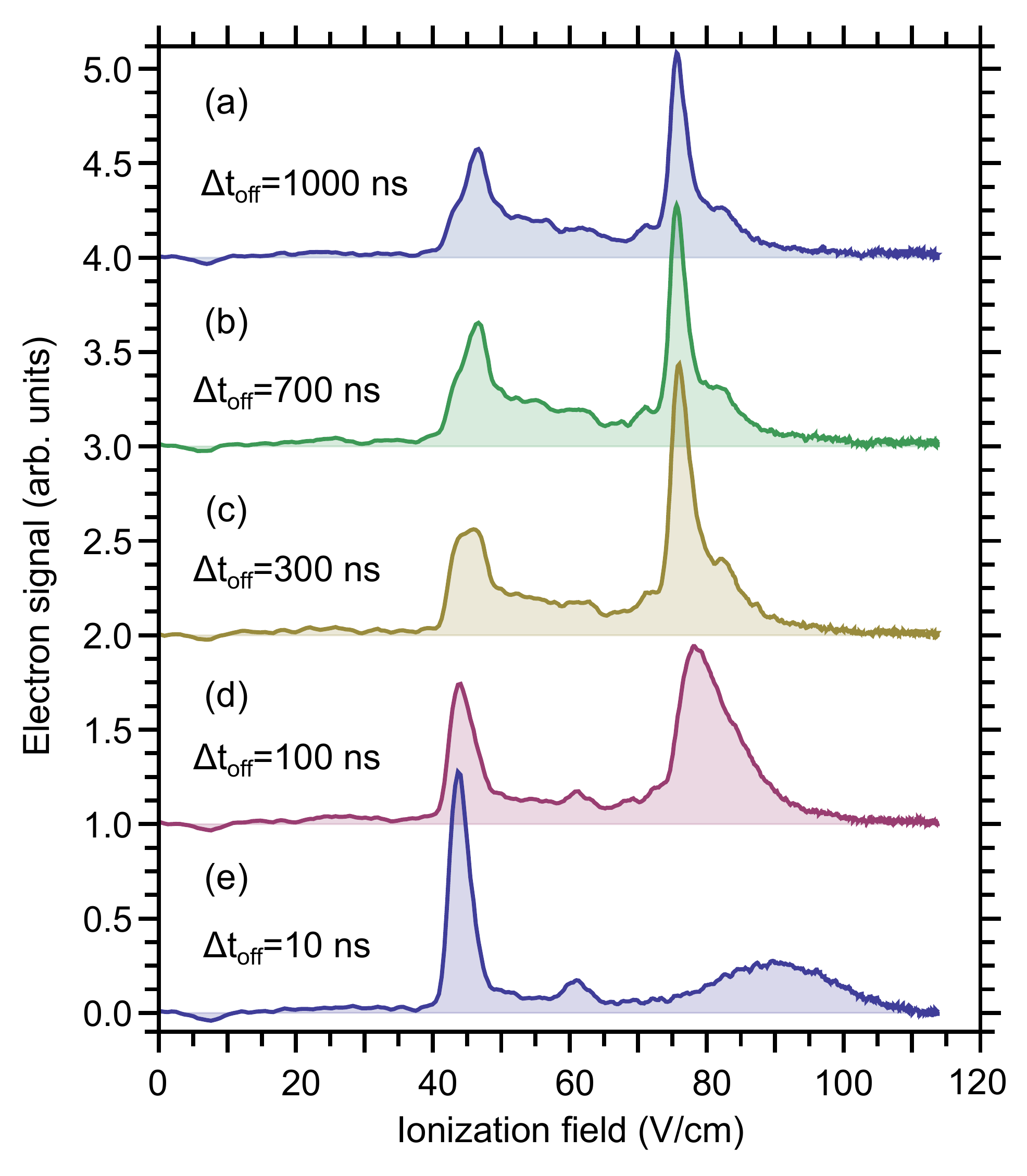}
\caption{Ramped electric field ionization measurements recorded after the excitation electric field was switched from $F_x=3.262$~V\,cm$^{-1}$ to zero in (a) 1000~ns, (b) 700~ns, (c) 300~ns, (d) 100~ns, and (e) 10~ns. The intensity maximum at an ionization field of $\sim75$~V\,cm$^{-1}$ in (a)-(c) corresponds to the $n=55$ circular state $|\mathrm{c}\rangle$.}
\label{fig4}
\end{figure}

The dependence of the circular state preparation efficiency on the rate at which the electric field is switched to zero after laser photoexcitation can be seen in the field ionization data presented in Fig.~\ref{fig4}. The datasets displayed in this figure represent the electron signal recorded at the MCP after ramped electric field ionization of the excited Rydberg atoms in the detection region [see Fig.~\ref{fig1}(a)]. The individual $n=55$ Rydberg-Stark states in helium ionize at a range of different electric fields. Adiabatic ionization~\cite{littman78} of the low-$|m_{\ell}|$ states occurs in fields of $\sim 40$~V\,cm$^{-1}$ while diabatic ionization of the outermost hydrogenic Stark state with a positive energy shift occurs at $\sim130$~V\,cm$^{-1}$~\cite{damburg83}. The scale on the horizontal axis in Fig.~\ref{fig4} results from a finite-element calculation of the electric field distribution in the ionization region of the apparatus for the measured time-dependent electrical potential applied to electrode $E_3$ for ionization.

For the measurements in Fig.~\ref{fig4}(e), in which the electric field was rapidly switched to zero within a time of $\Delta t_{\mathrm{off}}=10$~ns after the photoexcitation of the state $|\mathrm{c'}\rangle$ (i.e., when $\frac{\mathrm{d}\nu_{\mathrm{S}}}{\mathrm{d}t}\sim3.4\times10^{16}$~s$^{-2}$), the $n=55$ circular state is not efficiently prepared. This rapid switching of the excitation electric field to zero leads to a diabatic evolution of the excited state population into a wide range of $n=55$ sublevels. As a result electrons are detected over the entire range of ionization electric fields. The sharp peak at $\sim45$~V\,cm$^{-1}$ signifies adiabatic electric field ionization of the $m_{\ell}=0$ states while the broader feature at $\sim90$~V\,cm$^{-1}$ arises from diabatic ionization of the higher $|m_{\ell}|$ states.

When the atoms prepared in the circular state $|\mathrm{c}\rangle$ evolve adiabatically into the parallel electric and magnetic field region in which ionization occurs they ionize at a field of $\sim75$~V\,cm$^{-1}$~\cite{damburg79,damburg83}, approximately half way between the ionization field of the low-$|m_{\ell}|$ states and the outer positively-shifted Stark states. As can be seen in Fig.~\ref{fig4}(a)-(d), when the rate at which the electric field is switched off is reduced to $\frac{\mathrm{d}\nu_{\mathrm{S}}}{\mathrm{d}t}\simeq1.1\times10^{15}$~s$^{-2}$ [$\Delta t_{\mathrm{off}}=300$~ns, Fig.~\ref{fig4}(c)] a pronounced peak begins to appear in the electron signal at this ionization field. This peak gradually becomes narrower as $\frac{\mathrm{d}\nu_{\mathrm{S}}}{\mathrm{d}t}$ is reduced further to $\Delta t_{\mathrm{off}}=700$~ns and $\Delta t_{\mathrm{off}}=1000$~ns, and is the signature of efficient circular state production and selective ionization. The reduction in the amplitude of this feature for these largest values of $\Delta t_{\mathrm{off}}$ is attributed to a combination of the dependence of the electron collection efficiency on the ionization field strength and interactions of the strongly polarized ensemble of Ryd\-berg atoms with the electromagnetic field of the room temperature environment in which they are photoexcited during slow adiabatic transfer to the circular state. Further studies of the role of these interactions are planned in an apparatus cooled to cryogenic temperatures.

In contrast to experiments in which circular Rydberg states are prepared in the absence of external magnetic fields using the adiabatic microwave transfer method~\cite{hulet83}, the field ionization signal in parallel electric and magnetic fields is very sensitive to the evolution of the excited atoms into the ionization fields and the alignment of these fields. These effects have been studied in detail previously, in experiments involving the preparation of circular Rydberg states of rubidium with $n=67$~\cite{brecha93}, and indicate that misalignments of the `parallel' electric and magnetic fields in the ionization region on the order of $0.1^{\circ}$ result in the detection of a significant fraction of the initially prepared circular states as $m_{\ell}=0$ states. Such an observation is therefore not an indication of incomplete circular state production in the excitation region but rather a reflection of the electric field ionization process. The predominance of the electron signal corresponding to direct ionization of the circular state in Fig.~\ref{fig4}(a) suggests that the alignment of the fields in the ionization region in the apparatus used here is better than the $0.3^{\circ}$ achieved in Ref.~\cite{brecha93} for which the intensity of the electron signal associated with the $m_{\ell}=0$ states was greater than that from the circular state in all measurements presented.

\section{Microwave spectroscopy}\label{sec:MT}

\begin{figure}[!h]
\includegraphics[width=0.45\textwidth]{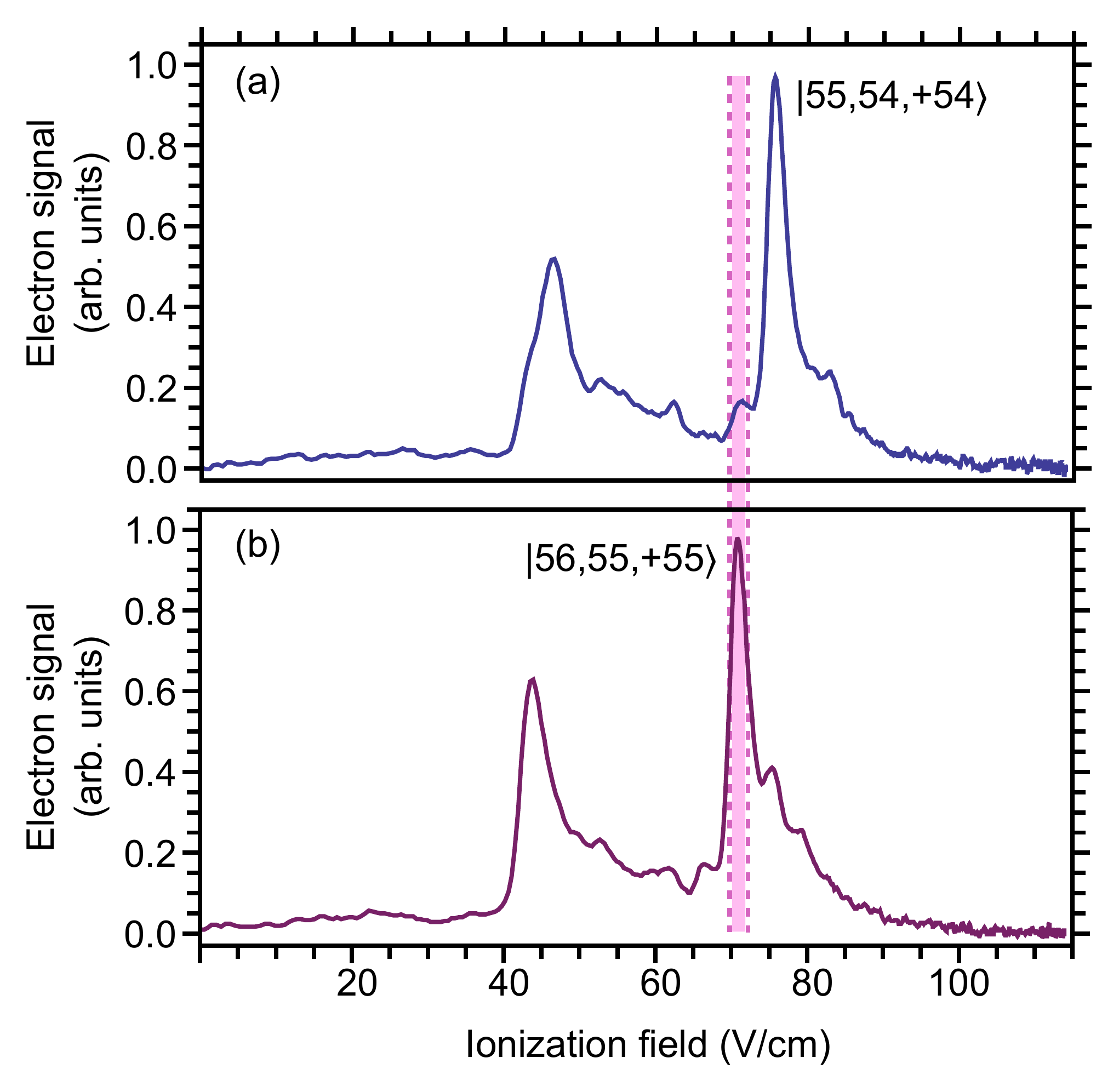}
\caption{Electric field ionization signal associated with (a) the $|55,54,+54\rangle$ state, and (b) the $|56,55,+55\rangle$ state. In (b) the population was transferred to the $|56,55,+55\rangle$ state by a resonant microwave pulse of 1~$\mu$s duration.}
\label{fig5}
\end{figure}

To verify the purity of the $n=55$ circular states prepared, microwave spectroscopy of the $|55,54,+54\rangle\rightarrow|56,55,+55\rangle$ circular-to-circular state transition was performed. In the absence of external fields, this transition occurs at 38.4890~GHz. However, the magnetic field of $\sim$15~G present in the microwave interaction region causes a Zeeman shift of the transition of $\Delta E_{\mathrm{Z}}/h\simeq20$~MHz. The transfer of population between the $n=55$ and $n=56$ circular states was identified from changes in the ionization electric field of the states. An example of the dependence of the electron signal at the MCP on the circular state populated is displayed in Fig.~\ref{fig5}. In Fig.~\ref{fig5}(a) the signal recorded following preparation of the $|55,54,+54\rangle$ circular state is presented. The electron signal recorded with a pulsed microwave field resonant with the $|55,54,+54\rangle\rightarrow|56,55,+55\rangle$ transition, and set to transfer all population to the latter is displayed in Fig.~\ref{fig5}(b). The difference in the direct ionization electric field of the two circular states is $\Delta F_{\mathrm{ion}}=4.75$~V\,cm$^{-1}$. From this data it can be seen that the electron signal resulting from direct ionization of each of the states can be completely separated. The corresponding integration window used in recording subsequent microwave spectra is indicated by the shaded region between the dashed vertical lines in this figure. 

As discussed in Section~III, the combined electric and magnetic fields in the detection region of the apparatus lead to the circular state ionizing through a range of sublevels within the same $n$-manifold. Consequently, upon applying the microwave pulse to address the $|55,54,+54\rangle\rightarrow|56,55,+55\rangle$ transition all of the features contributing to the ionization profile must shift to lower fields when the $|56,55,+55\rangle$ state is populated. This effect is clearly seen by comparing Fig.~\ref{fig5}(a) and (b), and further validates the interpretation of the ionization profiles discussed above and in Ref.~\cite{brecha93}.

\begin{figure}[!h]
\includegraphics[width=0.4\textwidth]{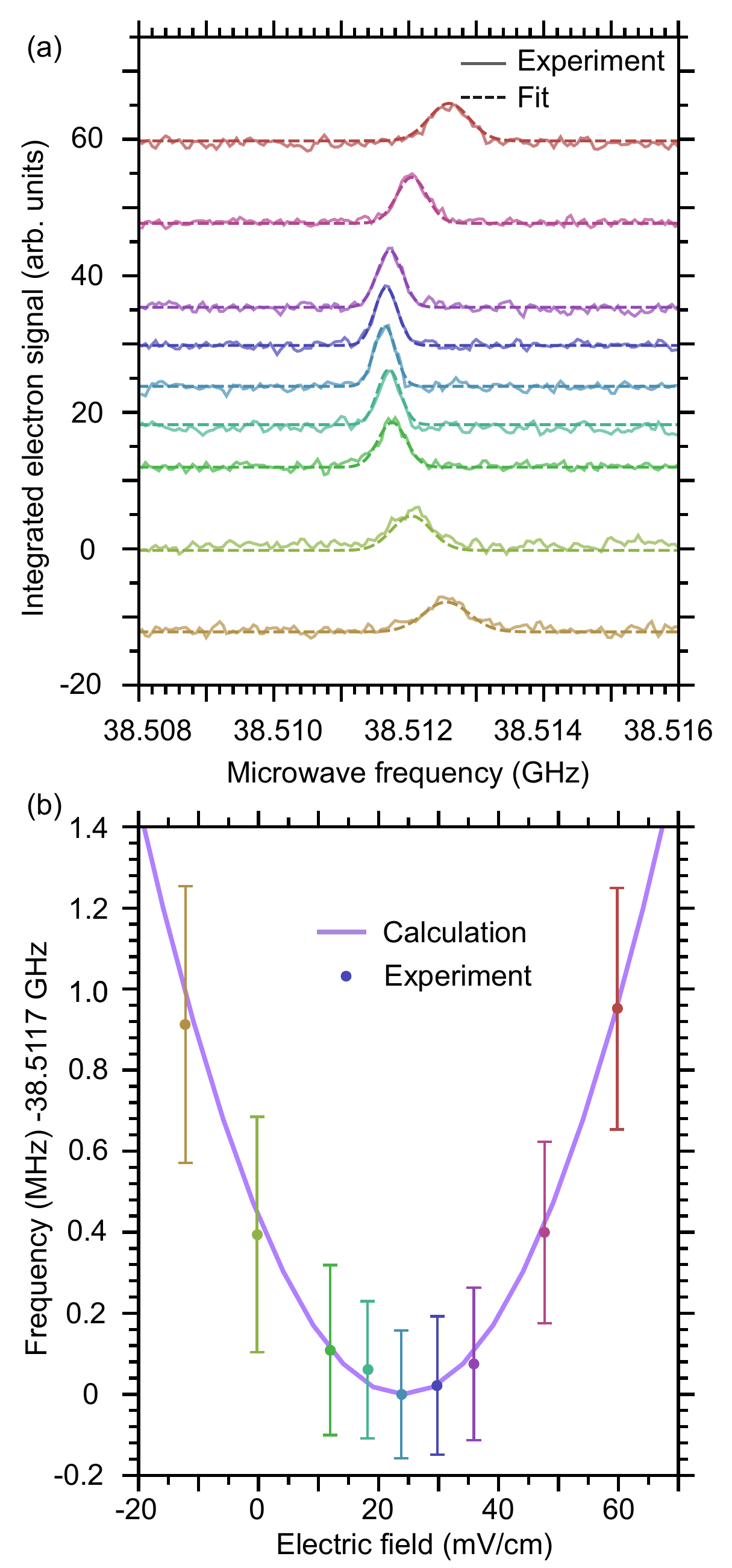}
\caption{(a) Microwave spectra of the $|55,54,+54\rangle\rightarrow|56,55,+55\rangle$ transition in applied electric fields ranging from $F_x=-12$ to $+60$~mV\,cm$^{-1}$ as indicated by the vertical offset in each case. Gaussian functions fitted to each experimentally recorded spectrum are indicated by the dashed curves. (b) Dependence of the resonant microwave transition frequency in (a) on the field $F_x$ (points), together with the calculated dependence of the transition frequency (continuous curve). The error bars on the experimental data represent one standard deviation of the Gaussian function in each spectrum in (a).}
\label{fig6}
\end{figure}

Because the Rydberg atoms in the experiment travel perpendicular to the applied magnetic field, they experience an induced motional electric field $\vec{F}_{\mathrm{mot}}=\vec{v}\times\vec{B}$. This gives rise to a motional Stark effect~\cite{crosswhite79,elliot95}. Since the mean velocity of the atoms in the beam is $\sim2000$~m\,s$^{-1}$ in the experiment, and the perpendicular magnetic field was $B_z\simeq15$~G, the induced motional electric field is $\sim30$~mV\,cm$^{-1}$. To determine the precise magnitude of, and cancel, this field, microwave spectra of the $|55,54,+54\rangle\rightarrow|56,55,+55\rangle$ transition in electric fields ranging from $F_x=-12$ to $+60$~mV\,cm$^{-1}$ were recorded. For each of these measurements microwave pulses of 3~$\mu$s duration were applied. The resulting spectra obtained by integrating the electron signal within the shaded region in Fig.~\ref{fig5}(b) are displayed in Fig.~\ref{fig6}(a). As the electric field was adjusted, the resonant transition frequency changed, reflecting the quadratic Stark shifts of the $|55,54,+54\rangle$ and $|56,55,+55\rangle$ states. In each spectrum the resonance frequency was determined by fitting a Gaussian function to the experimental data [dashed curves in Fig.~\ref{fig6}(a)]. The dependence of the microwave transition frequency on the applied electric field was then extracted [points in Fig.~\ref{fig6}(b)] and compared to the calculated dependence of the transition frequency on the electric field strength [continuous curve in Fig.~\ref{fig6}(b)]. The error bars on the experimental data in Fig.~\ref{fig6}(b) correspond to one standard deviation of the fitted Gaussian functions in Fig.~\ref{fig6}(a). These data indicate that the motional Stark effect is cancelled when an offset electric field of $F_x=+24$~mV\,cm$^{-1}$ is applied.

The spectra in Fig.~\ref{fig6}(a) were recorded at low microwave intensity such that on resonance $\sim$10\% of the population was transferred from the $|55,54,+54\rangle$ state to the $|56,55,+55\rangle$ state. As can be seen in Fig.~\ref{fig6}(a), the measured linewidth of the transition is narrowest when the cancellation field is applied. The spectral width of the transition increases when the electric field deviates from this cancellation field because the states become slightly polarized and therefore more sensitive to electric field inhomogeneities and electrical noise~\cite{zhelyazkova15a,zhelyazkova15b}. From the transition frequency in the presence of this cancellation electric field, the magnetic field in the excitation region could be determined to be $B_z=15.776$~G.

\begin{figure}[!h]
\includegraphics[width=0.5\textwidth]{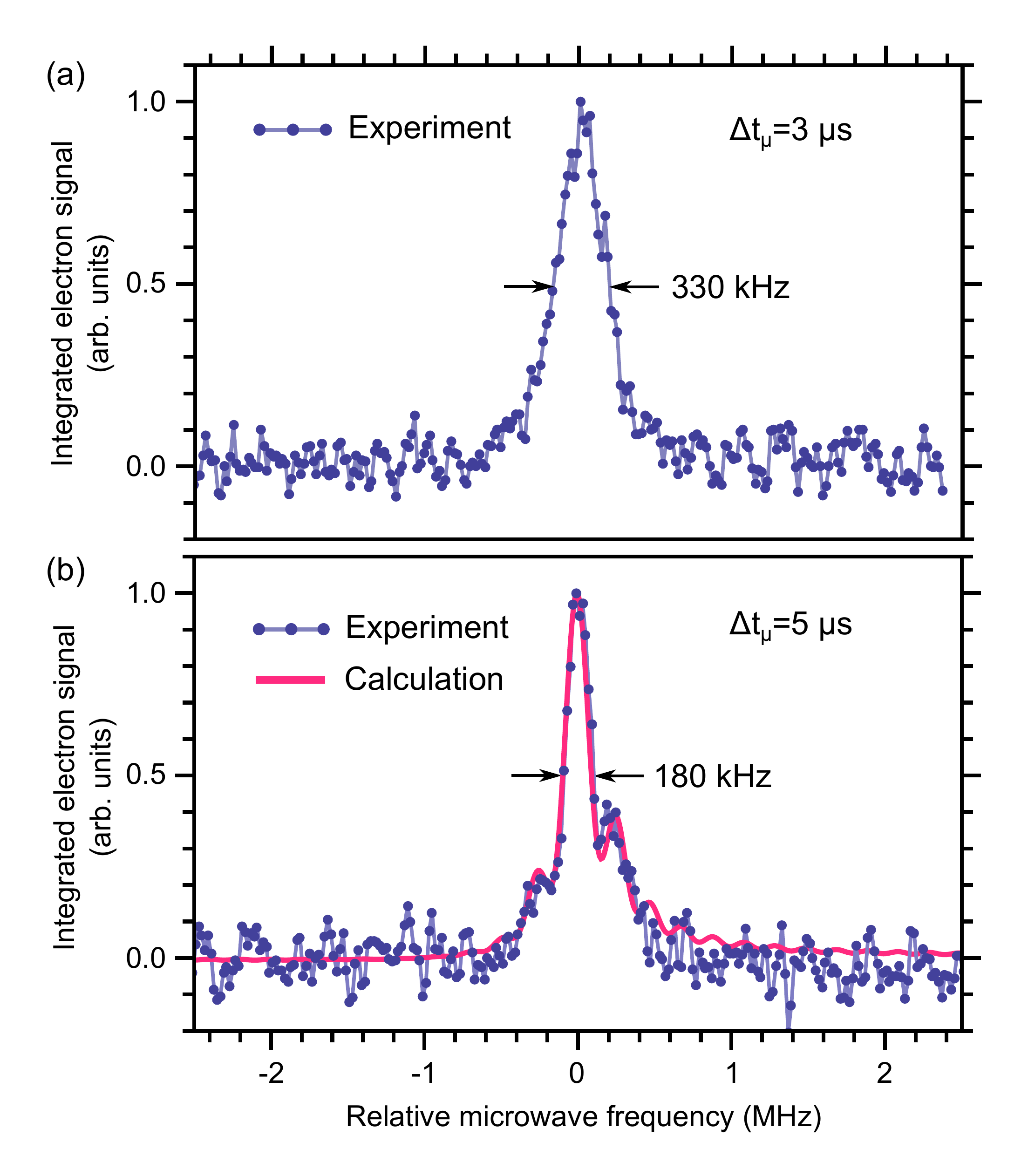}
\caption{Microwave spectra of the $|55,54,+54\rangle\rightarrow|56,55,+55\rangle$ transition (dots) for (a) 3~$\mu$s and (b) 5~$\mu$s long microwave pulses. The microwave frequency on the horizontal axis is offset by $-38.5117$~GHz. Also shown in (b) is a calculated Rabi resonance lineshape for a frequency dependence of the microwave intensity in the interaction region of $\sim3.3\times10^{-12}$~W\,cm$^{-2}$\,MHz$^{-1}$ (pink continuous curve).}
\label{fig7}
\end{figure}

To obtain more detailed information on the coherence of the $|55,54,+54\rangle\rightarrow|56,55,+55\rangle$ microwave transition and the spectral resolution in this experimental configuration, two further sets of measurements were performed. Microwave spectra were recorded of the transition at a microwave intensity of $I_{\mathrm{\mu}}\sim4\times10^{-12}$~W\,cm$^{-2}$, and for pulse durations of 3~$\mu$s and 5~$\mu$s. These spectra are presented in Fig.~\ref{fig7} (a) and (b), respectively. The measured spectral feature in Fig.~\ref{fig7}(a) has an approximately Gaussian profile with a FWHM of 330~kHz. Extending the microwave pulse duration to $5~\mu$s yields a more highly resolved spectral feature that exhibits a Rabi resonance lineshape~\cite{allen87}. The central peak of the spectral profile in Fig.~\ref{fig7}(b) has a FWHM of 180~kHz. The sidebands on either side of this central peak are characteristic of the Rabi resonance lineshape, however, the asymmetry in the intensity of these sidebands indicates a frequency dependence of the microwave intensity in the interaction region in the apparatus. A simulation of the lineshape, in which a linear frequency dependence of the microwave intensity, and the convolution with a Gaussian spectral profile with a FWHM of 130~kHz, were included resulted in the calculated spectral profile (continuous curve) overlaid on the experimental data in Fig.~\ref{fig7}(b). The good qualitative agreement between the result of this calculation and the experimental data indicates that in the apparatus the microwave intensity changes by $\sim3.3\times10^{-12}$~W\,cm$^{-2}$\,MHz$^{-1}$ across the spectral feature. Such frequency-dependent changes in the microwave intensity could be reduced in the future by improving the coupling of the microwave radiation into the apparatus.

\begin{figure}[!h]
\includegraphics[width=0.5\textwidth]{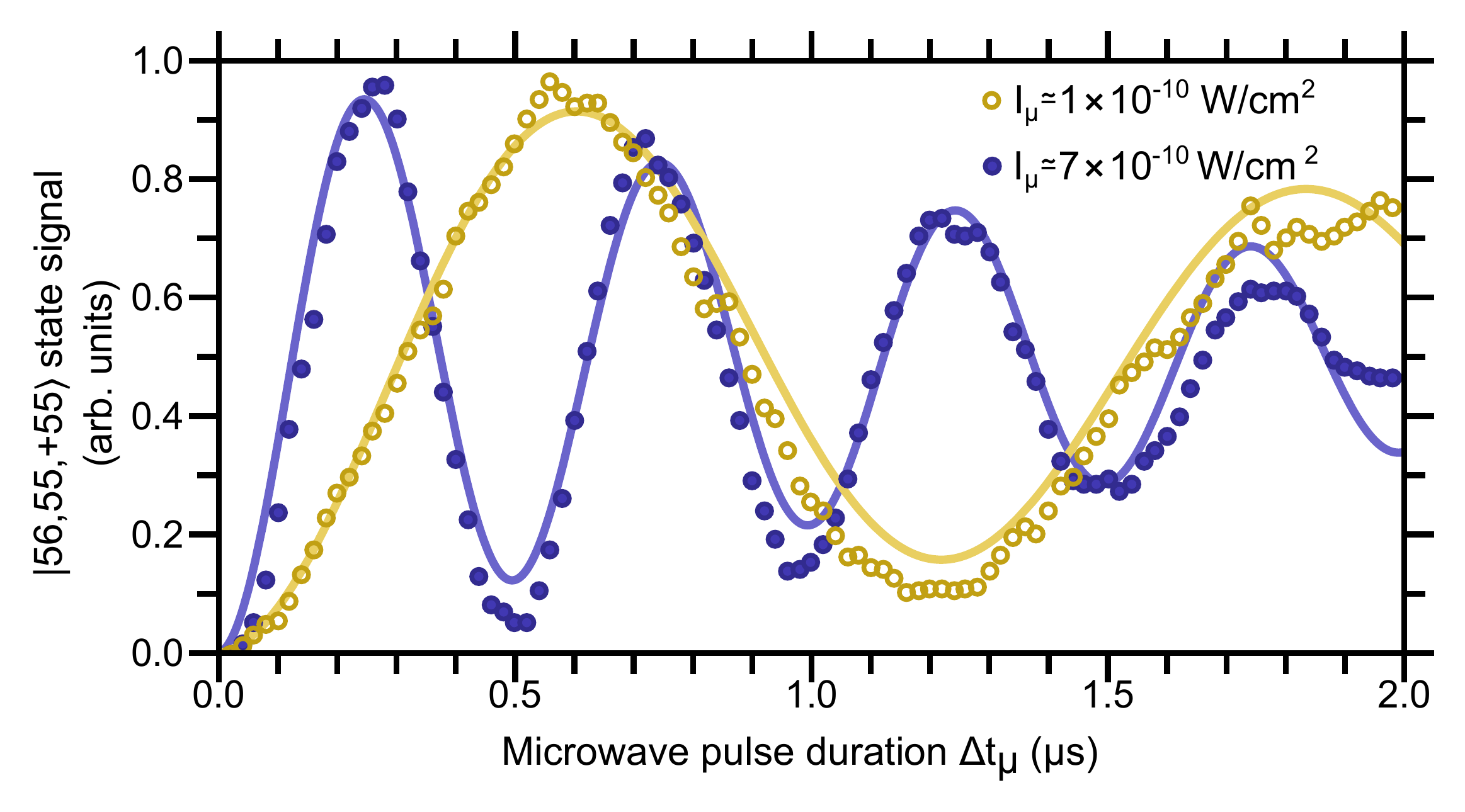}
\caption{Oscillations in the fraction of the atoms in the $|56,55,+55\rangle$ state for microwave intensities of $I_{\mathrm{\mu}}\simeq1\times10^{-10}$~W\,cm$^{-2}$ (open yellow points) and $I_{\mathrm{\mu}}\simeq7\times10^{-10}$~W\,cm$^{-2}$ (filled purple points) as a function of the microwave pulse duration. For each measurement the microwave frequency was held constant at $38.5117$~GHz. An exponentially decaying sinusoidal function fitted to each set of data (continuous curves) yields a coherence time of 1.76~$\mu$s (3.58~$\mu$s) when $I_{\mathrm{\mu}}\simeq7\times10^{-10}$~W\,cm$^{-2}$ ($I_{\mathrm{\mu}}\simeq1\times10^{-10}$~W\,cm$^{-2}$).}
\label{fig8}
\end{figure}

The coherence of the $|55,54,+54\rangle\rightarrow|56,55,+55\rangle$ transition in the present apparatus was determined from a set of measurements in which the duration of the pulse of resonant microwave radiation (38.5117~GHz) was adjusted between zero and 2~$\mu$s, in steps of 0.02~$\mu$s. At each step, the population in the $|56,55,+55\rangle$ state was detected. The evolution of this fraction of the atoms is displayed in Fig.~\ref{fig8} for two different microwave intensities, $I_{\mathrm{\mu}}\simeq1\times10^{-10}$~W\,cm$^{-2}$ (open yellow points), and $I_{\mathrm{\mu}}\simeq7\times10^{-10}$~W\,cm$^{-2}$ (filled purple points). As the duration of the microwave pulse was extended, Rabi oscillations are observed which gradually decohere. The coherence time of these oscillations is seen to depend significantly on the intensity of the microwave field, being shorter for higher intensities. Fitting exponentially decaying sinusoidal functions (continuous curves in Fig.~\ref{fig8}) to the experimental data leads to coherence times of 1.76~$\mu$s when $I_{\mathrm{\mu}}\simeq7\times10^{-10}$~W\,cm$^{-2}$, and 3.58~$\mu$s when $I_{\mathrm{\mu}}\simeq1\times10^{-10}$~W\,cm$^{-2}$. The long radiative lifetimes of the $|55,54,+54\rangle$ and $|56,55,+55\rangle$ states ($\sim46$~ms) do not contribute to decoherence. Instead, the transit of the Rydberg atoms through the microwave interaction region with its electric, magnetic and microwave field inhomogeneities dominates the decoherence observed in the experiment. However, the dependence of the coherence time on the microwave intensity suggests that off-resonant coupling to nearby states must also play a role even at these very low microwave intensities. In future experiments with microwave fields propagating in two-dimensional coplanar waveguides and confined in resonators, it will therefore be important to employ low microwave intensities, and further reduce the sensitivity of the atoms to stray electric fields by increasing and stabilizing the magnetic field strength in the appropriate regions of the apparatus to minimize decoherence.     

\section{Conclusion}
We have demonstrated efficient production of circular Rydberg states of helium with $n=55$ using the crossed electric and magnetic fields method. This approach is particularly effective in the triplet Rydberg states owing to the significant $\ell=0$ character of the outermost $m_{\ell}=0$ Stark states, required for circular state preparation, at the Inglis-Teller limit. Helium atoms in these excited states are ideally suited to probing atom-surface interactions, and coupling to chip-based microwave resonators, avoiding the detrimental effects of alkali metal adsorbates~\cite{hattermann12}.\\

\begin{acknowledgments}
This work was supported financially by the Department of Physics and Astronomy and the Faculty of Mathematical and Physical Sciences at University College London, and the Engineering and Physical Sciences Research Council under Grant No.~EP/L019620/1.
\end{acknowledgments}

\end{document}